\documentclass{article}

\usepackage{arxiv}

\usepackage[T1]{fontenc}    
\usepackage{hyperref}       
\usepackage{url}            
\usepackage{booktabs}       
\usepackage{amsfonts}       
\usepackage{nicefrac}       
\usepackage{microtype}      
\usepackage{lipsum}
\usepackage{graphicx}
\usepackage{multirow}
\usepackage{amsmath}
\usepackage{subcaption}
\usepackage{biblatex}
\usepackage{hyperref}
\usepackage{xurl}

\graphicspath{ {./images/} }

\title{The optimization of exact multi-target quantum search algorithm based on MindSpore}

\author{
 Shijin Zhong \\
  Anhui Engineering Research Center of Vehicle Display Integrated Systems, School of Integrated Circuits\\
   Anhui Polytechnic University\\
   Wuhu, China 241000 \\
  \texttt{zsj1079477926@163.com} \\
   \And
 Wei Li \\
  Anhui Engineering Research Center of Vehicle Display Integrated Systems, School of Integrated Circuits\\
   Anhui Polytechnic University\\
   Wuhu, China 241000 \\
  \texttt{2698291619@qq.com} \\
  \And
 Guangzhen Dai \\
  Anhui Engineering Research Center of Vehicle Display Integrated Systems, School of Integrated Circuits\\
   Anhui Polytechnic University\\
   Wuhu, China 241000 \\
  \texttt{daigzh@ahpu.edu.cn} \\
  \And
Daohua Wu \\
  Anhui Engineering Research Center of Vehicle Display Integrated Systems, School of Integrated Circuits\\
   Anhui Polytechnic University\\
   Wuhu, China 241000 \\
  \texttt{daohuawu@ahpu.edu.cn} \\
}

\begin{document}
\maketitle
{\begin{abstract}
Grover’s search algorithm has attracted great attention due to its quadratic speedup over classical algorithms in unsorted database search problems. However, Grover’s algorithm is inefficient in multi-target search problems, except in the case of 1/4 of the data in the database satisfying the search conditions. Long presented a modified version of Grover’s search algorithm by introducing a phase-matching condition, which can search for the target state with zero theoretical failure rate. In this work, we present an optimized exact multi-target search algorithm based on the modified Grover’s algorithm, by transforming the canonical diffusion operator to a more efficient diffusion operator, which can solve the multi-target search problem with a 100$\%$ success rate while requiring fewer gate counts and shallower circuit depth. After that, the optimized multi-target algorithm for four different items, including 2-qubit with 2 targets, 5-qubit with 2 targets, 5-qubit with 4 targets, and 6-qubit with 3 targets, are implemented on MindSpore framework. The experimental results show that, compared with Grover’s algorithm and the modified Grover's algorithm, the proposed algorithm can reduce the quantum gate count by at least 21.1$\%$ and the depth of the quantum circuit by at least 11.4$\%$ and maintain a 100$\%$ success probability.

Our code are available at \url{https://github.com/mindsporelab/models/tree/master/research/arxiv_papers/GROVER-OP.} Thanks for the support provided by MindSpore Community.
\end{abstract}}


\section{Introduction}
Quantum computation, due to its great advantage of speed over classical computation, and has attracted great attention over past decades \cite{bib1,bib2,bib3}. 
Although quantum computers are still in the research stage, some of the related research work such as quantum algorithms \cite{bib4,bib5,bib6}, have made amazing achievements both in 
theory and experiment. Wherein Grover's search algorithm is considered as one of the most popular quantum algorithms \cite{bib7}. The original Grover's search 
algorithm was proposed in 1996, which mainly solved the problem of searching for a target element in an unordered database with $N={2}^{n}$ elements \cite{bib6}. 
The classical algorithms usually search for the target element with the traversal search method in the database, which needs an average of $N/2$ times to find the 
target element, and its query complexity can be expressed as $O(N)$. Grover's algorithm, compared with classical algorithms, using the quantum properties of quantum coherence 
and superposition, provides a quadratic speedup by amplifying the probability amplitude of the target element, its query complexity is $O(\sqrt{N})$. 
Due to its great advantages in unordered database searching, Grover's algorithm leads to tremendous potential applications in the fields of information and 
computation \cite{bib8,bib9,bib10,bib11}. However, Grover's algorithm is a probabilistic algorithm, which means it can not search for the target element accurately except in the case 
of target number $M=N/4$. Especially, when $M=N/2$, the Grover's algorithm 
does not work. To solve this shortcoming, lots of modified versions of Grover's 
algorithm were proposed \cite{bib12,bib13,bib14,bib15,bib16}. Wherein the modified version proposed by Long \cite{bib13} is the most representative work, in which a phase-matching condition was introduced to construct a exact algorithm. For a search problem with a target element $|\tau\rangle$ in an unordered database, with $N={2}^{n}$ elements, the main idea of the modified Grover's algorithm can 
be carried out as \cite{bib17}: Replace the two $180^\circ$ phase inversion by phase rotation angle $\phi=2\arcsin(\sin[\pi/(4J+6)]/\sin\beta)$, where $J$ and $\beta$  satisfy $J\geq J_{min}=\left\lfloor(\pi/2-\beta)/2\beta\right\rfloor$ and $\beta=\arcsin(\sqrt{1/N})$. 
Here, $J$ can be chosen to be any integer equal to or larger than ${J}_{min}$. After $J$+1 times of iteration, the quantum state of the database can be expressed as $\left | \varphi   \right \rangle=e^{i[(\pi-\phi) / 2+J(\pi+\phi)]}|\tau\rangle$. Thus, the target state can be found out exactly. 
Later, the modified Grover's algorithm was strictly proved to be the simplest exact search algorithm \cite{bib18}.

Another key issue about quantum algorithms is resource estimation. Quantum algorithm is realized through the quantum circuit which is usually estimated by quantum gate count 
and quantum circuit depth. With the development of quantum technology, quantum computing has entered the era of noisy intermediate-scale quantum (NISQ), 
which attracts scholars' attention to resource estimation of quantum algorithms \cite{bib19,bib20,bib21,bib22,bib23,bib24}. In 2022, Zhou and Qiu et al. proposed an exact distributed Grover's algorithm by combining 
distributed computing and quantum computation \cite{bib22}. Its main idea is to divide a large-scale quantum computing task into multiple small-scale subtasks, and then implement the 
subtasks on each node parallelly by using quantum entanglement and quantum communication. By simulating on MindSpore framework, it was proved that their algorithm can search for a single-target exactly and requires fewer quantum gates and smaller circuit depth compared to the original Grover's algorithm. In 2023, Wu et al. proposed 
a circuit optimization method to optimize the block-level Oracle circuit which reduced the iterations and circuit depth \cite{bib23}. By combining the circuit optimization method with 
the divide-and-conquer idea, the authors presented a two-stage quantum search algorithm and simulated it on another quantum computing framework Cirq. The results show that, compared 
with Grover's algorithm, the circuit depth of their algorithm was reduced by 60$\%$. Recently, Kumar et al. optimized the quantum circuit of Grover's algorithm by transforming its diffusion 
operator to an efficient diffusion operator which resulted in a reduction of 40$\%$ and 32$\%$ for gate count and circuit depth, respectively \cite{bib24}.

Optimization of quantum algorithms and quantum circuits of multi-target search problems have also made great progress \cite{bib25,bib26,bib27,bib28}. Li et al. presented a quantum partial search algorithm 
for multi-target, in which the size of Oracles and time complexity were both smaller than counterparts of Grover's algorithm \cite{bib26}. Park et al. presented a quantum multi-programming of 
Grover's algorithm by applying quantum multi-programming to Grover's algorithm \cite{bib27}. Its main idea is to decompose Grover's algorithm using a partial diffusion operator and 
perform the decomposed circuits parallelly using quantum multi-programming. To test the validity of their Algorithm, the authors implemented it on several IBM quantum computers 
for one and three targets. The results proved that the success probability and circuit depth of their algorithm were better than the original Grover's algorithm and some other 
variations of Grover's algorithms. Gündüz et al. generalized Grover's algorithm to high-dimension multi-target Grover's search algorithm and successfully implemented it for 
two, three, and four targets on Cirq \cite{bib28}. The results showed their algorithm greatly reduced circuit complexity. Yet, their algorithm can't 
measure a multi-target state accurately. Up to now, there are few reports on an exact multi-target search algorithm that ensures 100$\%$ success in the search for more than 
a target (except for $M=N/4$), the gate count, and circuit depth are superior to Grover's algorithm.

Motivated by the discussions mentioned above, we present an optimized exact multi-target search algorithm that solves the multi-target search problem with with a 100$\%$ success rate, and requires fewer gate counts and shallower circuit depth compared to Grover’s algorithm and the modified Grover's algorithm. Specifically, (1) Expand the modified Grover's algorithm to a multi-target situation by introducing
a target number $M$ into the phase rotation angle, and construct a quantum circuit for the expanded multi-target search algorithm. (2) Optimize the quantum circuit by transforming the canonical diffusion operator to a more efficient diffusion operator that requires fewer gate counts and a shallower circuit depth. (3)  Implement the optimized quantum circuit of the multi-target algorithm for four different items, including 2-qubit with 2 targets, 5-qubit with 2 targets, 5-qubit with 4 targets, and 6-qubit with 3 targets, on MindSpore framework. The results show that, compared with Grover’s algorithm and the modified Grover's algorithm, the optimized algorithm can reduce the quantum gate count by at least 21.1$\%$ and the depth of the quantum circuit by at least 11.4$\%$ and maintain a 100$\%$ success probability.

The paper is organized as follows. In section 2, we give a brief overview of the original and modified Grover's algorithms. Subsequently, the optimization steps of the multi-target 
algorithm are introduced in section 3. Section 4 shows the implementation of an optimized exact multi-target algorithm, and the experimental results and discussion are presented 
in section 5. Finally, a brief conclusion is presented in section 6.

\section{Background}

In this section, we will briefly review the original Grover's algorithm \cite{bib6} and the modified Grover's algorithm \cite{bib13}.
\subsection{Original Grover's search algorithm}\label{subsec1}
Original Grover's search algorithm mainly solves the problem of searching for a single-target element in an unordered database. To understand Grover's algorithm easily, let us 
consider an unordered database with $N={2}^{n}$ elements. Define a Boolean function $f(x):\{0,1\}^n\rightarrow\{0,1\}$, where the search element $x \in\{0,1\}^n$. When the function satisfies $f(x=\tau)=1$, 
$x=\tau$ is the solution to the search problem, otherwise $f(x)=0$. The database can be described in a quantum system as a basis vector $|x\rangle$ $(x=1,2,3 \ldots \tau, \ldots N)$, where $\tau$  represents the 
target state. The purpose of Grover's algorithm is to identify the target state $|\tau\rangle$ with the highest probability. 

Specifically, the process of Grover's search algorithm can be fulfilled as the following steps:

\textbf{Step 1}: Initialize n qubits $|0\rangle^{\otimes n}$ on a quantum register, and obtain

\begin{equation}\label{eqn-1} 
	\left|\varphi_0\right\rangle=|0\rangle^{\otimes n}.
\end{equation}

\textbf{Step 2}: Apply the Hadamard gate to the quantum register, and prepare the initial state into a superposition state with equal probability amplitude

\begin{equation}\label{eqn-2} 
	|\varphi\rangle=H^{\otimes n}\left|\varphi_0\right\rangle=H^{\otimes n}|0\rangle^{\otimes n}=\frac{1}{\sqrt{2^n}}\sum_{x=0}^{2^n-1}{\left|x\right\rangle.}
\end{equation}

\textbf{Step 3}: Apply Oracle operator $ O_{G}=I-2|\tau\rangle\langle\tau| $ to recognize the target state, and reverse the phase of the target state to $\left|x\right\rangle\rightarrow\left(-1\right)^{f\left(x\right)}\left|x\right\rangle$ by $f(x)=1$. 
Then apply the diffusion operator $D_G=2\left|\varphi\right\rangle\langle\varphi|-I$ to increase the probability amplitude of the target state. Actually, the Oracle operator and diffusion operator can be defined as the Grover operator as

\begin{equation}\label{eqn-3} 
	G=D_{G} O_{G}=(2|\varphi\rangle\langle\varphi|-I)(I-2|\tau\rangle\langle\tau|).
\end{equation}

Grover operator $G$ refers to an iterative operation that results in getting closer to the target state. After the \emph{k} iteration of the Grover operator, the initial state can be expressed as

\begin{equation}\label{eqn-4} 
	G^k\left(\varphi\right)=\cos{\left(\left(2k+1\right)\theta\right)}\left|\tau^,\right\rangle+\sin{\left(\left(2k+1\right)\theta\right)\left|\tau\right\rangle},
\end{equation}
\noindent
where $\left|\tau^,\right\rangle$ represents non-target state. Thus, the probability of ﬁnding out the target state is

\begin{equation}\label{eqn-5} 
	{P=\sin}^2{\left(\left(2k+1\right)\theta\right)}.
\end{equation}

Obviously, as $[\left(2k+1\right)\theta]$ gets closer to $\pi /2$, namely, $k\approx\frac{\pi}{4\theta}-\frac{1}{2}$, the probability approaches 1. However, since $[\left(2k+1\right)\theta]$ may not be 
exactly $\pi /2$, the maximum probability $P_{max}$ of finding out the target state is usually not 100$\%$. Actually, it is strictly proved that Grover's search algorithm is unable to find out 
the target state exactly, except in the case of 1/4 of the data in the database satisfying the search conditions.

\subsection{The modified Grover's algorithm}
To solve the shortcomings mentioned above, Long proposed a phase-matching theory for quantum search algorithm, and constructed an exact search algorithm, which found the target state with a theoretical success probability of 100$\%$ \cite{bib13}. Different from Grover’s algorithm, Long introduced a phase rotation to replace the two phase-inversions in Grover’s algorithm, and let the two phase rotation angles equal which was required by the phase-matching condition for a database with the evenly distributed linear superposition state\cite{bib29}. Thus, after $J+1$ iterations, the target state can be obtained with certainty.

Speciﬁcally, the modified Grover's algorithm can be given as follows:

\textbf{Step 1}: Same as Grover's algorithm, initialize n qubits $\left|0\right\rangle^{\otimes n}$ on a quantum register and obtain

\begin{equation}\label{eqn-6} 
	{\left|\varphi_0\right\rangle=\left|0\right\rangle}^{\otimes n}.
\end{equation}

\textbf{Step 2}: Apply Hadamard gate to the quantum register, and prepare the initial state into a superposition state with equal probability amplitude

\begin{equation}\label{eqn-7} 
	\left|\varphi_1\right\rangle=H^{\otimes n}{\left|\varphi_0\right\rangle=H}^{\otimes n}\left|0\right\rangle^{\otimes n}=\frac{1}{\sqrt{2^n}}\sum_{x=0}^{2^n-1}{\left|x\right\rangle.}
\end{equation}

\textbf{Step 3}: Define iteration operator $L=-D_LO_L$ by modifying Oracle operator and diffusion operator to ${O_{L}}=I+(e^{i\phi}-1){|\tau\rangle\langle\tau|}$ 
and $D_{L}=I+(e^{i\phi}-1)|\varphi\rangle\langle\varphi|$, respectively. To satisfy the phase-matching condition, we make two-phase rotation angles equal and modify them as

\begin{equation}\label{eqn-8} 
	\phi=2\arcsin[\frac{\sin{\left(\frac{\pi}{4J+6}\right)}}{\sin\beta}],
\end{equation}
\noindent
where $J$ and $\beta$ satisfy  

\begin{equation}\label{eqn-9} 
	J\geq J_{min}=\left\lfloor\frac{\frac{\pi}{2}-\beta}{2\beta}\right\rfloor,
\end{equation}
\noindent
and

\begin{equation}\label{eqn-10} 
	\beta=\arcsin{\left(\sqrt{\frac{1}{N}}\right)},
\end{equation}
\noindent
here, $\left\lfloor A\right\rfloor$ represents the integer part of $\ J_{min}$, and $J$ can be chosen to be any integer equal or larger than $J_{min}$. Repeat L operator $J+1$ times, and obtain \cite{bib17}

\begin{equation}\label{eqn-11} 
	\left|\varphi\right \rangle=e^{i[(\pi-\phi) / 2]+J(\pi+\phi)}|\tau\rangle,
\end{equation}
Thus, the target state will be measured with a success probability of 100$\%$.

\section{Optimized exact multi-target search algorithm}
By borrowing the idea of optimization method in \cite{bib24} and the modified Grover’s algorithm \cite{bib13}, we present an optimized exact search algorithm for multiple target items. Compared to Grover’s algorithm and the modified Grover’s algorithm, the optimized algorithm requires fewer gate counts and shallower circuit depth, and maintains a 100$\%$ success rate. The optimization of the exact multi-target search algorithm can be realized in three steps.

First, extend the modified Grover's algorithm to multi-target situations. When the number of target states is $M$, using the same three steps given in subsection 2.2 and modifying the phase rotation angle by replacing $\beta$ in equation (\ref{eqn-10}) with

\begin{equation}\label{eqn-12} 
	\beta=\arcsin{(\sqrt{\frac MN})}.
\end{equation}

The M target states can be found with certainty which was strictly proved in \cite{bib29}.

\begin{figure}
\centering
\includegraphics[width=0.8\textwidth]{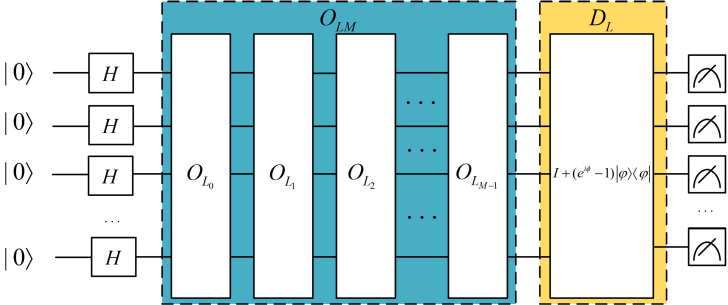}
\caption{Representation of quantum circuit for exact multi-target search algorithm.}
\label{fig1}
\end{figure}

Secondly, construct the quantum circuit for the multi-target search algorithm. Consider an unordered databased with $N=2^n$ element, $s_i\epsilon\left\{0,1\right\}^n\,(i=0, 1, 2….M-1)$ is the set of the targeted element. In order to identify $M$ target states, it is necessary to construct $M$ different Oracle operators which correspond to $M$ different target states. The serial arrangement of $M$ Oracle operators and one diffusion operator construct the multi-target search algorithm which is shown in figure \ref{fig1}.

The process of the exact multi-target search algorithm can be fulfilled as the following steps:

(1) The qubits initiated in the state $\left|\varphi_0\right\rangle=\left|0^{\otimes n}\right\rangle$ are transformed into an equal probability amplitude superposition state by the Hadamard gate.

\begin{equation}\label{eqn-13} 
	\left|\varphi_1\right\rangle=H^{\otimes n}\left|0\right\rangle^{\otimes n}=\frac{1}{\sqrt{2^n}}\sum_{x=0}^{2^n-1}{\left|x\right\rangle}.
\end{equation}

(2) The target states are searched out by repeating the $L_M$ operator $J+1$ time. Following that, the $L_M$ operator can be expressed as

\begin{equation}\label{eqn-14} 
	L_M=-D_LO_{LM}=\left[\begin{matrix}-e^{i\phi}(1+(e^{i\phi}-1){\sin}^2\beta)&-(e^{i\phi}-1)\sin\beta \cos\beta\\-e^{i\phi}(e^{i\phi}-1)\sin\beta \cos\beta&-e^{i\phi}+(e^{i\phi}-1){\sin}^2\beta\\\end{matrix}\right],
\end{equation}
\noindent
where

\begin{equation}\label{eqn-15} 
	O_{LM}=I+(e^{i\phi}-1)\sum_{i=1}^{M}|s_i\rangle\langle s_i|,
\end{equation}
\noindent
and

\begin{equation}\label{eqn-16} 
	D_L=I+(e^{i\phi}-1)|\varphi\rangle\langle\varphi|.
\end{equation}

Suppose $M$ is the number of target states, and $\left|s_i\right\rangle$ refers to the target elements. The $O_{LM}$ operator, at this point, consists of $M$ single Oracle operators arranged serially for identifying $M$ target states.
Let $\left|s_i\right\rangle=\left|x_{n-1},x_{n-2},\ldots,x_0\right\rangle$ be an n-qubit target state, the single Oracle operator can be realized using Pauli gate $X$ and Phase-shift gate $PS\left(\phi\right)$. 
Pauli gate $X$ is expressed in matrix form

\begin{equation}\label{eqn-17} 
	X=\left[\begin{matrix}0&1\\1&0\\\end{matrix}\right].
\end{equation}

In the target state $\left|x_i\right\rangle=\left|x_{n-1},x_{n-2},\ldots,x_0\right\rangle$, the gate $X_{g(x_i)}$ to $\left|x_i\right\rangle$ is implemented through equation (\ref{eqn-18}).

\begin{equation}\label{eqn-18} 
	g\left(x_i\right)=\left\{\begin{matrix}
		1,  & x_i=0\\ 
		0,  & x_i=1
	\end{matrix}\right..
\end{equation}

Phase-shift gate $PS\left(\phi\right)$ has the following matrix representation:

\begin{equation}\label{eqn-19} 
	PS\left(\phi\right)=\left[\begin{matrix}1&0\\0&e^{i\phi}\\\end{matrix}\right],
\end{equation}
\noindent
where $\phi$ is the phase rotation angle mentioned in equation (\ref{eqn-8}). Accordingly, the decomposition of the single Oracle operator $O_L$ operator is as follows:

\begin{equation}\label{eqn-20} 
	O_L=(\otimes_{\iota=0}^{n-1}X^{g(x_i)})C^nPS(\phi)(\otimes_{\iota=0}^{n-1}X^{g(x_i)}).
\end{equation}

\begin{figure}[ht]
	\centering
	\subfloat[]{
		\includegraphics[width=0.55\linewidth]{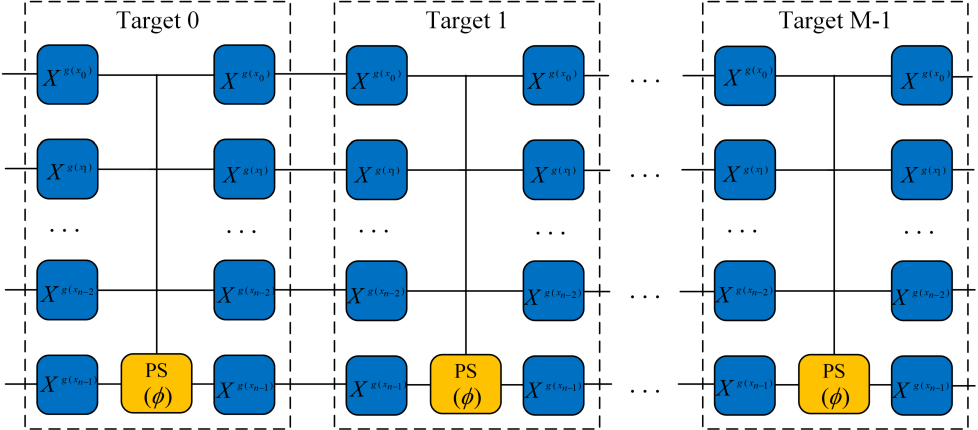}\label{fig2a}
	}\hfill
	\subfloat[]{
		\includegraphics[width=0.34\linewidth]{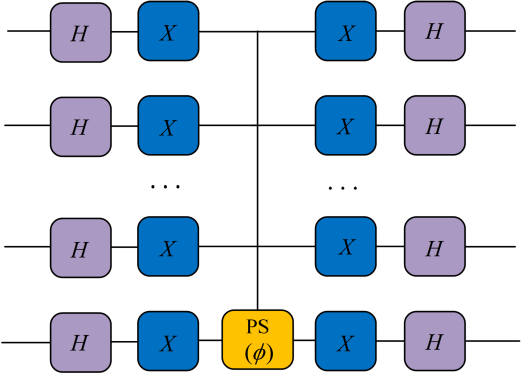}\label{fig2b}
	}
	\caption{Quantum circuit of Oracle operator and diffusion operation. (a) Oracle circuit for a multi-target search. (b) Diffusion circuit for n-qubit.}
	\label{fig2}
\end{figure}

Figure \ref{fig2}(\subref{fig2a}) shows the Oracle circuit for a multi-target search. $O_{LM}$ operator is followed by the $D_L$ operator. The equivalent of the $D_L$ diffusion operator in terms of unitary gates is described as:

\begin{equation}\label{eqn-21} 
	D_L=AC^nPS(\phi)A^{\dag}.
\end{equation}

The $A$ is a construction of $H$ gate and $X$ gate and the $A^{\dag}$ is a combination of $X$ gate and $H$ gate.The quantum circuit of the diffusion operator is shown in  figure \ref{fig2}(\subref{fig2b}).

(3) After repeating $L_M=-D_LO_{LM}$ on the database state $J+1$ times and then performing a measurement on the register, the target states will be obtained with certainty. 

\begin{figure}[ht]
	\centering
	\includegraphics[width=0.8\textwidth]{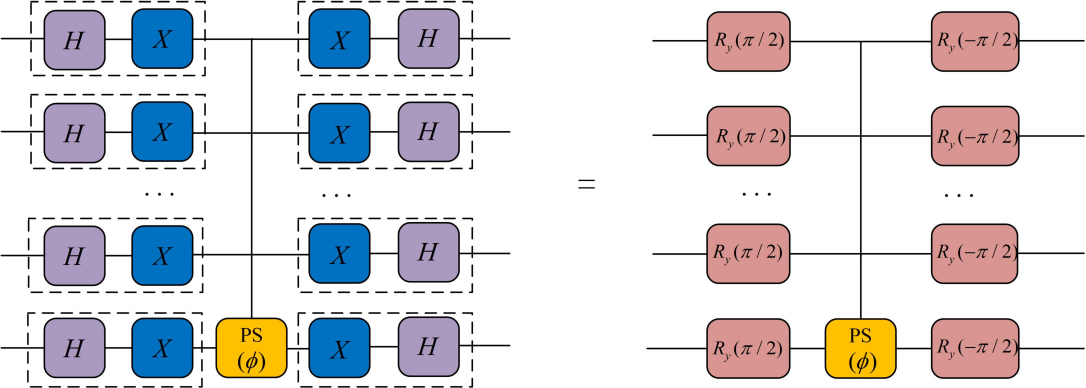}
	\caption{Merging of gates to transform canonical diffusion operator into optimized diffusion operator.}\label{fig3}
\end{figure}

Thirdly, optimization of the exact multi-target search algorithm. The novel algorithm is optimized by reducing the gate count from the quantum circuit of its diffusion 
operator, which in turn reduces the depth of the circuit.  The canonical diffusion 
operator revealed in figure \ref{fig2}(\subref{fig2b}) is represented as equation (\ref{eqn-21}). The iteration operator in the modified Grover's algorithm with multi-target is represented by $L_M=-AC^nPS(\phi)A^{\dag}O_{LM}$, where $O_{LM}$ consists of $M$ single Oracle operators arranged serially, $A=H^{\otimes n}X^{\otimes n}$ and $A^{\dag}=X^{\otimes n}H^{\otimes n}$. In this paper, the Hadamard gate $H$ is combined with the Pauli gate $X$ to construct a rotation gate $R_y(\pi/2)^{\otimes n}$. Similarly, the Pauli gate $X$ merged with the Hadamard gate $H$ to construct a rotation gate $R_y(-\pi/2)^{\otimes n}$. Thus, the diffusion operator can be transformed, as depicted in Figure \ref{fig3}.

The iteration operator of the optimized algorithm is represented by 

\begin{align}\label{eqn-22} 
	L_M&=-AC^nPS(\phi)A^{\dag}O_{LM}=-H^{\otimes n}X^{\otimes n}C^nPS(\phi)X^{\otimes n}{H}^{\otimes n}O_{LM}
	\notag
	\\&=-{R_y(\pi/2)}^{\otimes n}C^nPS(\phi){R_y(-\pi/2)}^{\otimes n}\cdot\sum_{i=1}^{M}(\otimes_{\iota=0}^{n-1}X^{g(x_i)})C^nPS(\phi)(\otimes_{\iota=0}^{n-1}X^{g(x_i)}).
\end{align}

\begin{figure}[ht]
	\centering
	\includegraphics[width=1\textwidth]{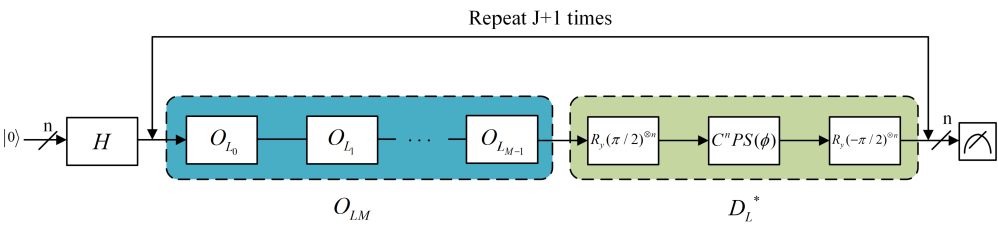}
	\caption{Schematic circuit of optimized exact multi-target search algorithm.}\label{fig4}
\end{figure}

Figure \ref{fig4} depicts the schematic circuit of the optimized exact multi-target search algorithm. 

\section{Implementation of the optimized exact multi-target search algorithm}
Due to the fact that a mature quantum computer has not yet been developed, a quantum computing framework becomes an important platform to implement, verify, and optimize quantum algorithms. 
MindSpore is one of the popular quantum computing frameworks that provides a quantum programming language to create, optimize, and run a quantum circuit for quantum algorithms. 
In this section, the optimized exact multi-target algorithm for 2-qubit with 2-target, 5-qubit with 2-target, 5-qubit with 4-target and 6-qubit with 3-target, are implemented 
on MindSpore framework, respectively.

\begin{figure}[ht]
	\centering
	\includegraphics[width=1\textwidth,trim=0 0 0 0,clip]{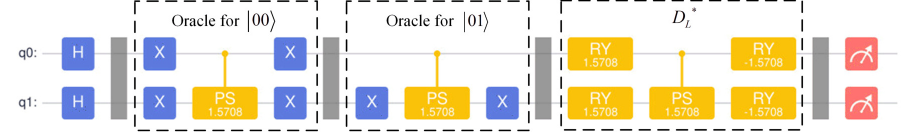}
	\caption{Circuit of 2-qubit the optimized exact multi-target search algorithm for two targets such that $\left|00\right\rangle$ and $\left|01\right\rangle$ states on the MindSpore framework.}\label{fig5}
\end{figure}

\begin{figure}[ht]
		\centering
		\includegraphics[width=0.5\textwidth,trim=0 0 0 40,clip]{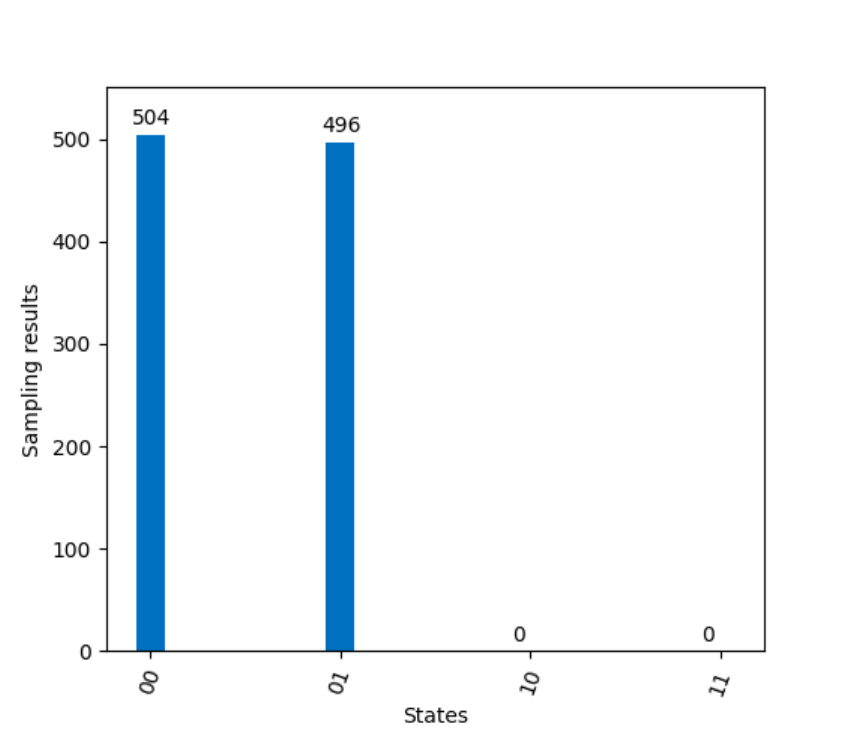}
		\caption{Implementation result of 2-qubit the optimized exact multi-target search algorithm for two targets such that $\left|00\right\rangle$ and $\left|01\right\rangle$ states on the MindSpore framework.}
		\label{fig6}
\end{figure}

In the case of 2-qubit with 2 targets. Consider an ${N=2}^2$ database and two random targets ($\left|00\right\rangle$ and $\left|01\right\rangle$) in the database ($\left|00\right\rangle$, $\left|01\right\rangle$, $\left|10\right\rangle$ and $\left|11\right\rangle$), each element in the database has an equal probability. Based on the discussions mentioned above, the quantum circuit of 2-qubit with 2 targets optimized algorithm is shown in figure \ref{fig5}.
The two dashed boxes in the circuit correspond to the Oracle operations of the two target states and the two Oracle operators with serial arrangement. The two-pair rotation gates $Ry(\pi/2)$ and $Ry(-\pi/2)$ are used in the diffusion operator according to equation (\ref{eqn-22}) for the last dashed boxes. The two Oracle operators and the optimized diffusion operator are both repeated 
for $J+1$ times for the detection of two target states with 100$\%$. The parameters in the quantum circuit are calculated as: $\phi=2\arcsin\left(\sin[\pi/(4J+6)]/\sin\beta\right)\approx1.5708$, 
$\beta=\arcsin{(\sqrt{2/2^{2}})}$ and $J=\left\lfloor\frac{\pi}{4} (\sqrt{2/2^{2}})-\frac{1}{2}\right\rfloor=0$, thus, the number of iterations is $J+1=1$. By 
sampling 10,00 times of the circuit depicted in figure \ref{fig5}, the sampling result is presented in figure \ref{fig6}.

\begin{figure}[ht]
	\centering
	\includegraphics[width=0.7\textwidth]{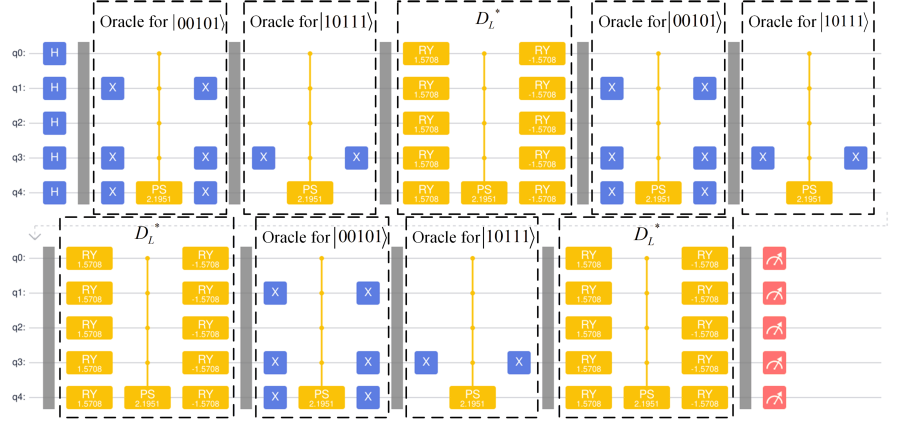}
	\caption{Circuit of 5-qubit the optimized exact multi-target search algorithm for two targets such that $\left|00101\right\rangle$ and $\left|10111\right\rangle$ states on the MindSpore framework.}\label{fig7}
\end{figure}

\begin{figure}[ht]
	\centering
	\includegraphics[width=0.55\textwidth,trim=0 0 0 40,clip]{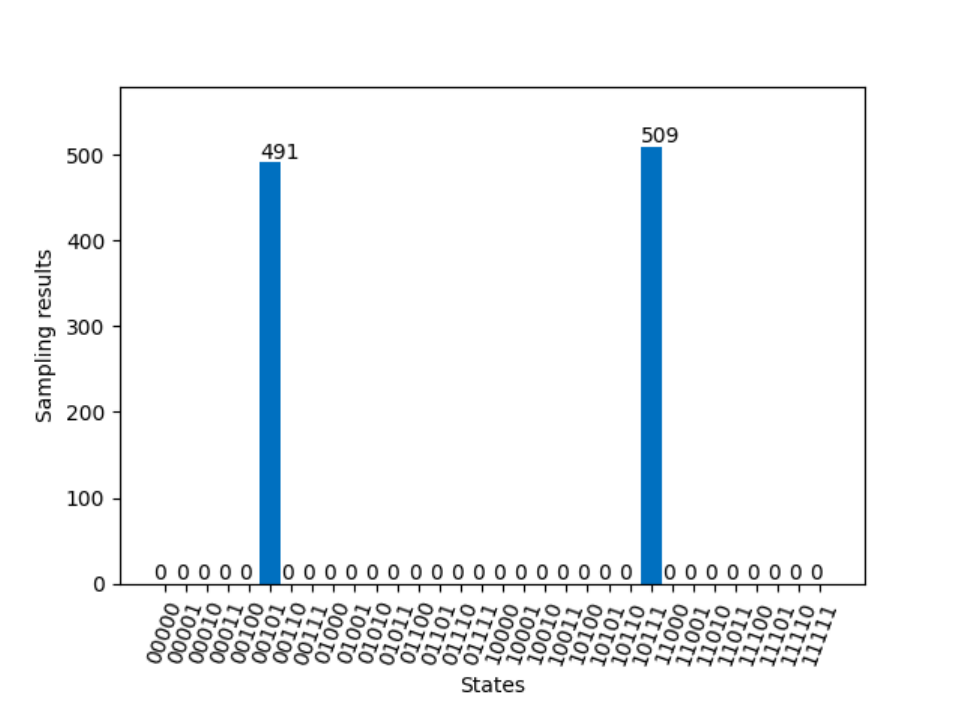}
	\caption{Implementation result of 5-qubit the optimized exact multi-target search algorithm for two targets such that $\left|00101\right\rangle$ and $\left|10111\right\rangle$ states on the MindSpore framework.}\label{fig8}
\end{figure}

In the case of 5-qubit with 2 targets. Consider an ${N=2}^5$ database with two random targets ($\left|00101\right\rangle$ and $\left|10111\right\rangle$). 
The quantum circuit is constructed in figure \ref{fig7}. The two dashed boxes in the circuit correspond to the Oracle operations of the two target states and the two Oracle operators with serial arrangement. The five-pair rotation gates $Ry(\pi/2)$ and $Ry(-\pi/2)$ are 
used in the diffusion operator for the last dashed boxes. The Oracles and the optimized diffusion operator are both repeated for $J+1$ times for the detection of two target states with 100$\%$. 
The parameters of the circuit are  $\phi=2\arcsin\left(\sin[\pi/(4J+6)]/\sin\beta\right)\approx2.1951$, $\beta=\arcsin{(\sqrt{2/2^{5}})}$ and $J=\left\lfloor\frac{\pi}{4} (\sqrt{2/2^{5}})-\frac{1}{2}\right\rfloor=2$, and the iteration number is $J+1=3$. By sampling 10,00 times of the quantum circuit, the implemented result is presented in figure \ref{fig8}.

\begin{figure}[ht]
	\centering
	\includegraphics[width=0.7\textwidth,trim=0 0 0 0,clip]{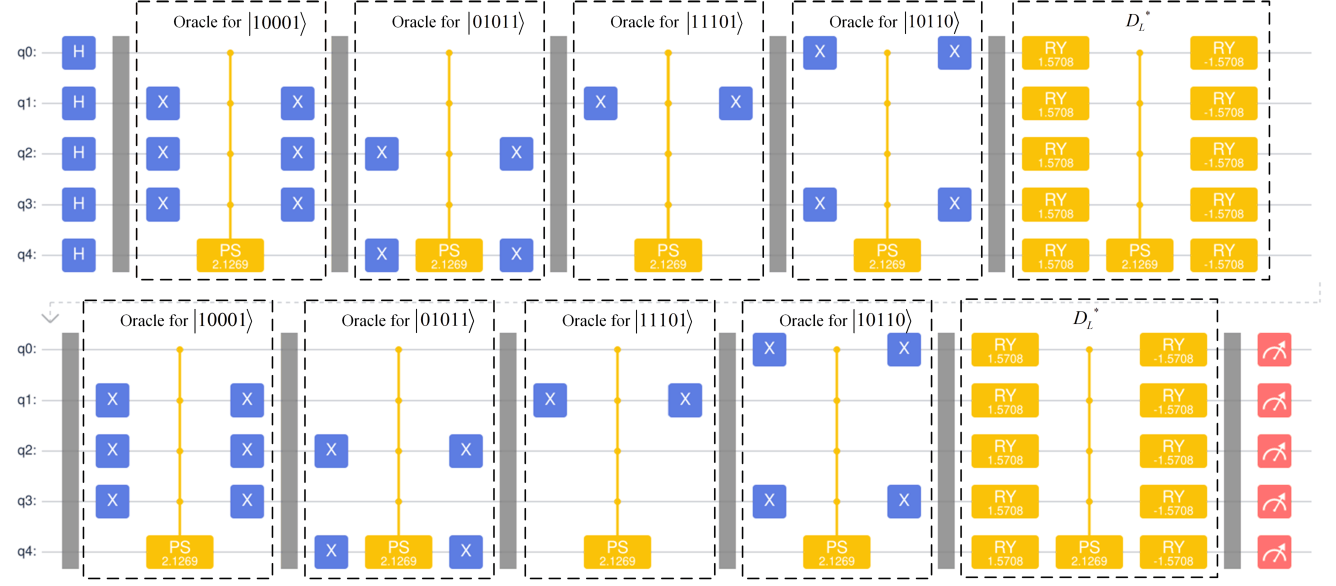}
	\caption{Circuit of 5-qubit optimized exact multi-target search algorithm for four targets such that $\left|10001\right\rangle$, $\left|01011\right\rangle$, $\left|11101\right\rangle$, and $\left|10110\right\rangle$ states on the MindSpore framework.}\label{fig9}
\end{figure}

\begin{figure}[ht]
	\centering
	\includegraphics[width=0.6\textwidth,trim=0 0 0 35,clip]{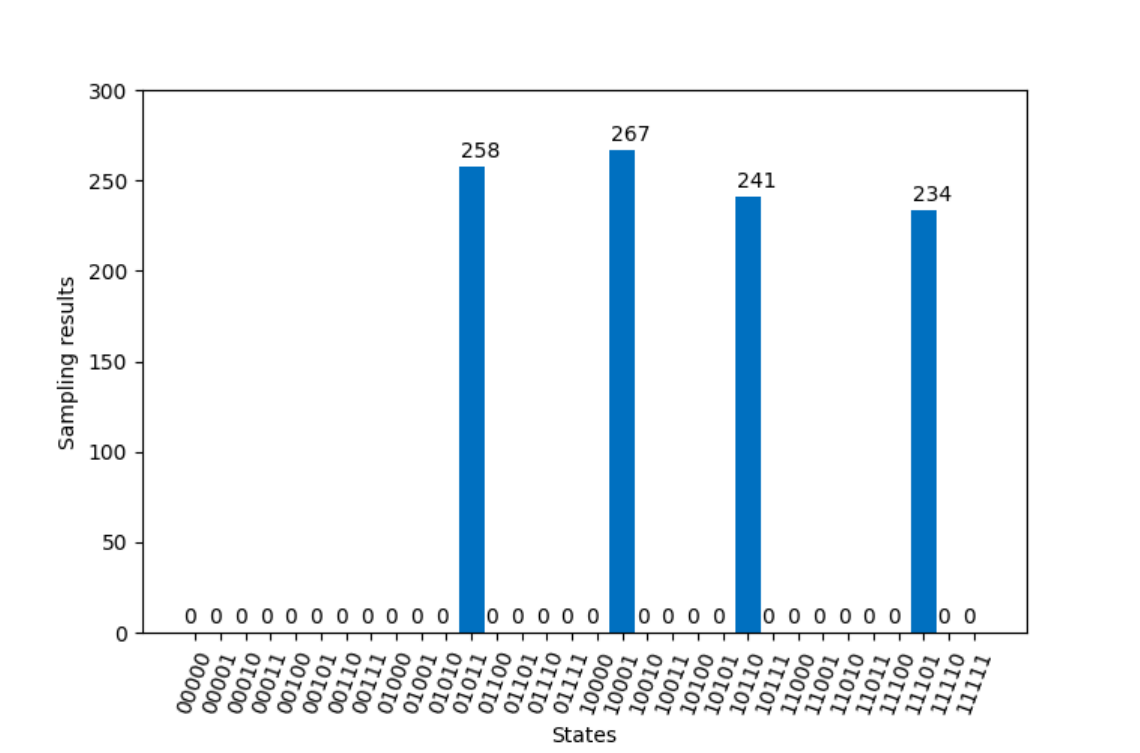}
	\caption{Implementation result of 5-qubit the optimized exact multi-target search algorithm for four targets such that $\left|10001\right\rangle$, $\left|01011\right\rangle$, $\left|11101\right\rangle$, and $\left|10110\right\rangle$ states on the MindSpore framework.}\label{fig10}
\end{figure}

For the case of 5-qubit with 4 targets. Consider a ${N=2}^5$ database with four random targets ($\left|0001\right\rangle$, $\left|1011\right\rangle$, $\left|1101\right\rangle$ and $\left|0110\right\rangle$). 
The quantum circuit is constructed in figure \ref{fig9}. Four different Oracle operators with serial arrangement are constructed to recognize the four target states and five-pair 
rotation gates $Ry(\pi/2)$ and $Ry(-\pi/2)$ are used in the diffusion operator. Different dashed boxes represent different operators in the circuit, the Oracle and the optimized diffusion operator are both repeated for $J+1$ times for the detection 
of four target states with 100$\%$. The parameters of the circuit are $\phi=2\arcsin\left(\sin[\pi/(4J+6)]/\sin\beta\right)\approx2.1269$, $\beta=\arcsin{(\sqrt{4/2^{5}})}$ 
and $J=\left\lfloor\frac{\pi}{4} (\sqrt{4/2^{5}})-\frac{1}{2}\right\rfloor=1$, the number of iterations is $J+1=2$. By sampling 10,00 times of the quantum circuit, the implementation outcome is presented in figure \ref{fig10}.

\begin{figure}[ht]
	\centering
	\includegraphics[width=0.7\textwidth,trim=0 0 0 0,clip]{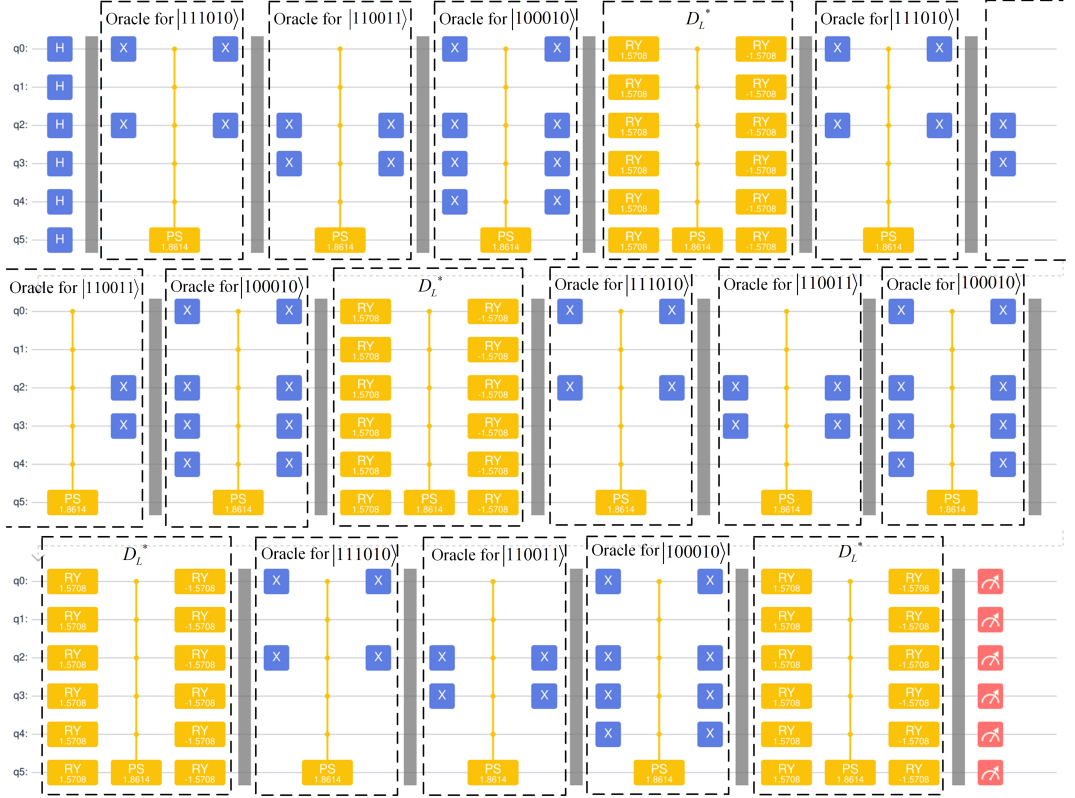}
	\caption{Circuit of 6-qubit optimized exact multi-target search algorithm for three targets such that $\left|100010\right\rangle$, $\left|110011\right\rangle$ and $\left|111010\right\rangle$ states on the MindSpore framework.}\label{fig11}
\end{figure}

\begin{figure}[ht]
	\centering
	\includegraphics[width=1\textwidth,trim=0 0 0 35,clip]{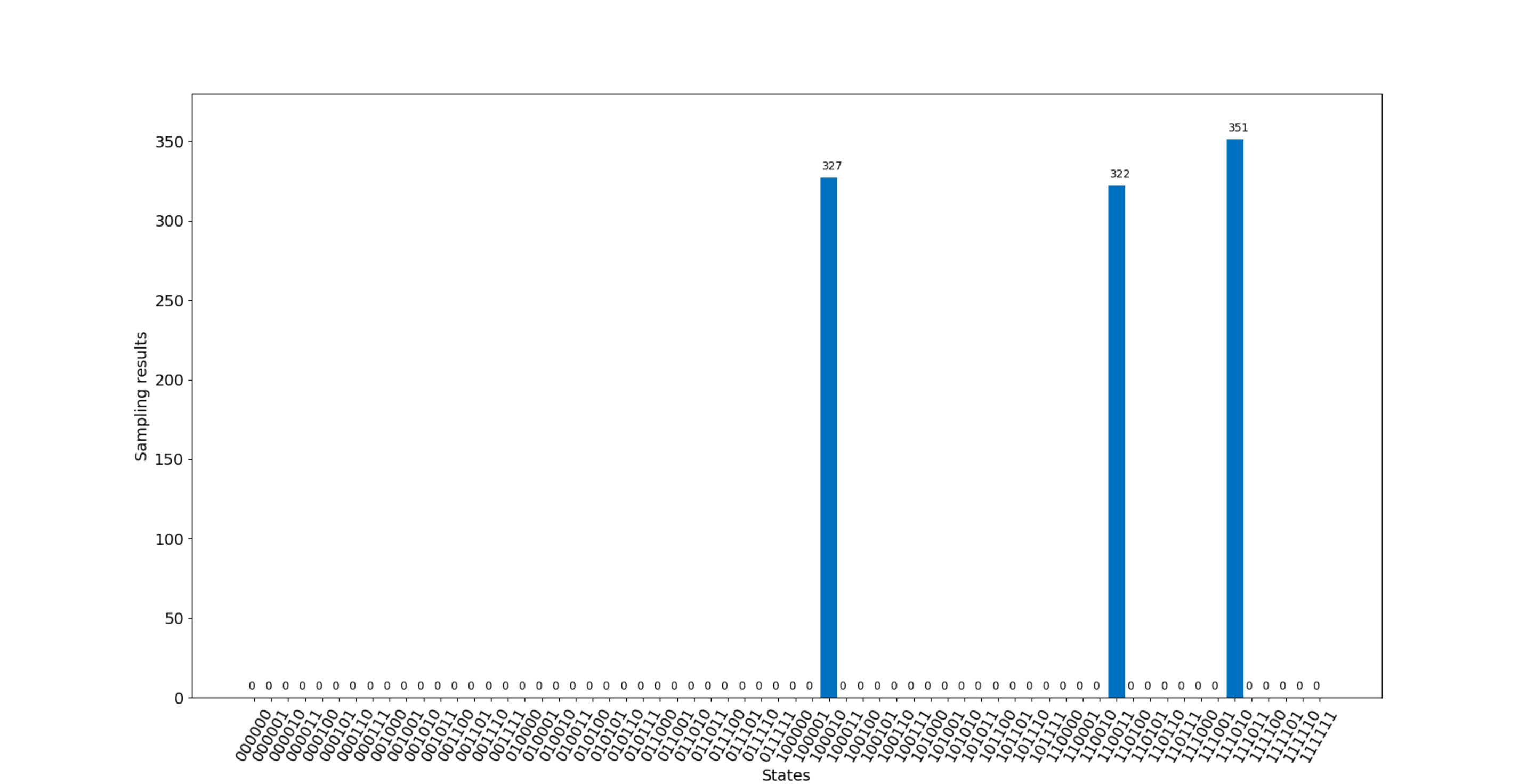}
	\caption{Implementation result of 6-qubit the optimized exact multi-target search algorithm for three targets such that $\left|100010\right\rangle$, $\left|110011\right\rangle$ and $\left|111010\right\rangle$ states on the MindSpore framework.}\label{fig12}
\end{figure}

For the case of 6-qubit with 3 targets. Consider a ${N=2}^6$ database with three random targets ($\left|100010\right\rangle$, $\left|110011\right\rangle$ and $\left|111010\right\rangle$). The quantum circuit is constructed as figure \ref{fig11}. Three different Oracle operators with serial arrangement are constructed to recognize the three target states and six-pair rotation gates $Ry(\pi/2)$ and $Ry(-\pi/2)$ are used in the diffusion operator. The Oracle and the diffusion operator are both repeated for $J+1$ times for the detection of three target states with 100$\%$. The parameters of the quantum circuit are $\phi=2\arcsin\left(\sin[\pi/(4J+6)]/\sin\beta\right)\approx1.8614$, $\beta=\arcsin{(\sqrt{3/2^{6}})}$
and $J=\left\lfloor\frac{\pi}{4} (\sqrt{3/2^{6}})-\frac{1}{2}\right\rfloor=3$, the number of iterations is $J+1=4$. By sampling 10,00 times of the quantum circuit, the implement result is presented in figure \ref{fig12}.

\section{Experimental results and discussion}\label{sec5}

In this section, experimental results, including success probability, quantum gate count and circuit depth of the quantum circuit are discussed. After that, comparisons 
of the proposed optimized algorithm against Grover's algorithm and the modified Grover's algorithm are presented.

\subsection{Success probability}\label{subsec3}
Success probability is an important parameter, which assesses the accuracy of quantum search algorithms. Figure \ref{fig6} demonstrates the sampling result 
for the case of 2-qubit with 2 targets item. By sampling 1000 times of the quantum circuit, the random target state $\left|00\right\rangle$ is detected 
504 times, and $\left|01\right\rangle$ is detected 496 times, the other non-target states are not yet detected. The result indicates that the total success 
probability of searching for $\left|00\right\rangle$ and $\left|01\right\rangle$ is 100$\%$. The same results can be obtained for other situations from figure \ref{fig8}, figure \ref{fig10} and figure \ref{fig12}, respectively. Table~\ref{tab1} gives a comparison of success probability for the original Grover's algorithm, the modified Grover's algorithm and our optimized algorithm. 
As can be seen in Table~\ref{tab1}, the novel optimized algorithm, the same as the modified Grover's algorithm, maintains a 100$\%$ success probability and does not change with the number of 
targets and the size of the database, which is obviously superior to Grover's algorithm. Especially, for the case of a 2-qubit with 2 targets situation, which corresponds 
to the target number $M=N/2$, the success probability of Grover's algorithm is 49.5$\%$, and that means Grover's algorithm fails. While the proposed optimized algorithm is 
capable of achieving an exact search in this case.

\begin{table*}[ht]
	\renewcommand{\arraystretch}{2} 
	\resizebox{\linewidth}{!}{
		\begin{tabular}{@{}llllll}
			\hline
			Qubit number & Target number & Targets              & Grover's algorithm & The modified Grover's algorithm  & Optimized algorithm \\ \hline
			2            & 2             & 00,01       & 49.5$\%$          & 100$\%$ &100$\%$         \\
			5            & 2             & 00101,10111    & 95.8$\%$    & 100$\%$    & 100$\%$         \\
			5            & 4             & 01011,10001,10110,11101    & 94.7$\%$  & 100$\%$  & 100$\%$   \\
			6            & 3             & 100010,110011,111010    & 96.3$\%$  & 100$\%$  & 100$\%$   \\
			\hline
		\end{tabular}
	}
	\caption{Comparison of the success probability for the three different algorithms based on the MindSpore framework.}\label{tab1}
\end{table*}

\subsection{Quantum gate count}

\begin{figure}[ht]
	\centering
	\includegraphics[width=0.6\textwidth,trim=0 0 0 10,clip]{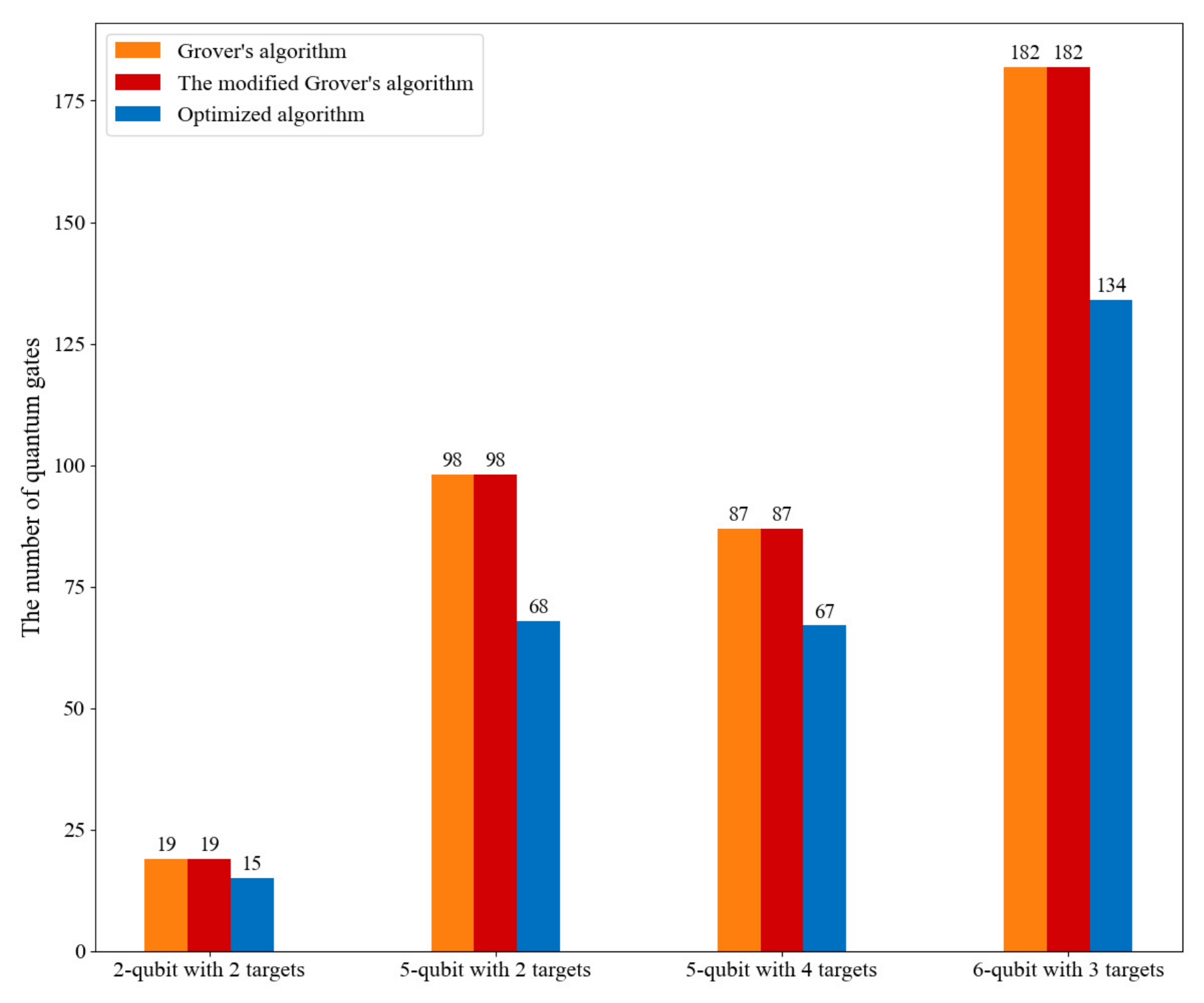}
	\caption{Comparison of the quantum gate count for the three different algorithms based on the MindSpore framework.}\label{fig13}
\end{figure}

Quantum gate is the fundamental unit of a quantum circuit, which rapidly increases with the size of the quantum circuit. 
With the increase in target number and database size, the implementation of the quantum algorithm needs a 
large-scale quantum circuit, which makes the quantum circuit become complex and leads to various noises or 
errors in the quantum circuit. In this paper, the proposed multi-target search algorithm is optimized by replacing 
the combination of Hadamard gate $H$ with Pauli gate $X$ of canonical diffusion operator with a rotation gate $Ry(\theta)$ 
for each target qubit. The total quantum gate count is apparently reduced for the proposed optimized multi-target search 
algorithm, which makes the quantum circuit more compact compared with Grover's algorithm and the modified Grover's algorithm. To understand 
the advantage of the proposed optimized multi-target search algorithm directly, the total gate count of Grover's, modified Grover's and 
proposed optimized algorithms for the above four experimental items are shown in figure \ref{fig13}. The experimental results in figure \ref{fig13} show 
that the proposed optimized algorithm contains a lesser number of quantum gates than Grover's algorithm and the modified Grover's algorithm. 
Taking the case of 5-qubit with 2 targets as an example, the total number of quantum gates of Grover’s algorithm, the modified Grover's algorithm, and the proposed optimized algorithm are 98, 98, and 68, respectively. It is obviously observed a reduction of 30.6$\%$ in quantum gate count compared to Grover’s algorithm and the modified Grover's algorithm. Similarly, the reductions of 21.1$\%$, 23.0$\%$ and 26.4$\%$ can be observed from the other three experimental situations. The reduction of the quantum gate count can effectively decrease the complexity of quantum circuits, which makes it possible to implement quantum algorithms more easily for large-scale quantum computing in the future.

\begin{figure}[ht]
	\centering
	\includegraphics[width=0.8\textwidth,trim=0 0 0 0,clip]{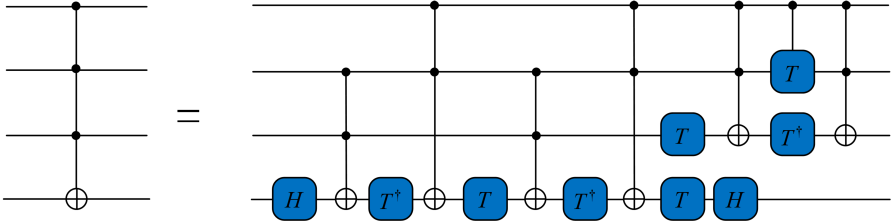}
	\caption{Quantum circuit for the decomposition of $C^3X$ gate.}\label{fig14}
\end{figure}

To compare the gate count between different circuits directly, the quantum circuit of the optimized algorithm, the modified Grover's algorithm, and Grover's algorithm are decomposed and compared against each other. By employing the decomposition method in \cite{bib31}, the multi-qubit $C^nX$ in Grover's algorithm and $C^nPS$ gates in the modified Grover's algorithm and optimized algorithm are decomposed into a combination of single-qubit and double-qubit gates. For example, the $C^3X$ can be decomposed into the quantum circuit in figure \ref{fig14}, which contains six $C^2X$ gates, one $CT$ gate, two $H$ gates, and six $T$ gates. The $CT$ gate is equivalent to the $CPS(\pi/4)$ gate and $T$ gate is equivalent to the $PS(\pi/4)$ gate. For the $C^nX$ gate, it can be decomposed using the same method. 

\begin{figure}[ht]
	\centering
	\includegraphics[width=0.8\textwidth,trim=0 0 0 0,clip]{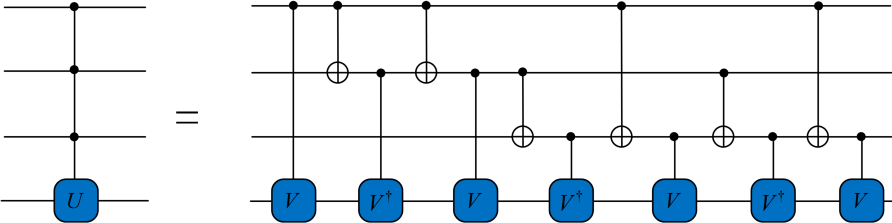}
	\caption{Quantum circuit for the decomposition of $C^3U$ gate.}\label{fig15}
\end{figure}

For the $C^3PS$ gate, it can be decomposed into the quantum circuit in figure \ref{fig15}, which contains six $CNOT$ gates and seven $CV$ gates. The $U$ represents another single-qubit gate that fulfills the condition $V^4 = U$. If $U = PS(\theta)$, then $V = PS(\theta/4)$ and $V^\dag = PS(-\theta/4)$ can be readily derived.

The decomposed quantum circuits of Grover’s algorithm, the modified Grover's algorithm, and the proposed optimized algorithm illustrated in Table~\ref{tab2} are compared on the basis of gates and gate count. Take the 5-qubit with 2 targets as an example, the number of quantum gates of Grover’s algorithm, the modified Grover's algorithm and the proposed optimized algorithm are 5552, 2006 and 1976, respectively. It is obviously observed a reduction of 63.9$\%$ compared with Grover’s algorithm and 1.5$\%$ compared with the modified Grover's algorithm, respectively.

\begin{table*}[ht]
	\renewcommand{\arraystretch}{1.7} 
	\resizebox{\linewidth}{!}{
		\begin{tabular}{@{}llllllllll}
			\hline
			Quantum algorithm & Qubits and targets & $H$ & $X$ & $T$ & $R_y$ & $CPS$ & $CT$ & $CNOT$  & Number of gates   \\ \hline
			Grover’s algorithm &  & 809  & 54  &  2646  & 0  & 27  & 54 & 1962  & 5552      \\
			The modified Grover's algorithm & \multirow{1}{*}{5-qubit with 2 targets}    & 287 &  54  & 864  & 0  & 81 & 18  & 702 & 2006   \\
			optimized algorithm   &   & 257  & 24  & 864   & 30  & 81 & 18 & 702  & 1976            \\
			\hline
			Grover’s algorithm &    & 895  & 62  & 2940   & 0  & 30 & 60 & 2180  & 6167         \\
			The modified Grover's algorithm & \multirow{1}{*}{5-qubit with 4 targets} & 305  & 52  &  960  & 0  & 90  & 20 & 780  & 2207      \\
			optimized algorithm &  & 285  & 32  & 960 & 20  & 90 & 20 & 780    &2187        \\
			\hline
			Grover’s algorithm &    & 7822  & 112  &26550   & 0  & 375 & 540 & 19710  & 55109        \\
			The modified Grover's algorithm & \multirow{1}{*}{6-qubit with 3 targets} & 3411  & 568  &  10234  & 0  & 272  & 224 &8824  & 23533      \\
			optimized algorithm &  & 3363  & 520  & 10234 & 48  & 272 & 224 & 8824   &23485        \\
			\hline
	\end{tabular}
	}
	\caption{Comparison of the quantum gate count for the three different algorithms after decomposition based on the MindSpore framework.}\label{tab2}
\end{table*}

\subsection{Circuit depth}

\begin{figure}[ht]
	\centering
	\includegraphics[width=0.8\textwidth,trim=0 0 0 0,clip]{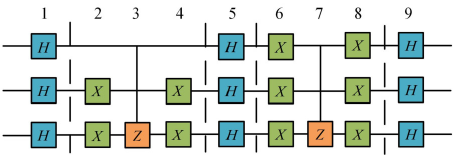}
	\caption{A quantum circuit of Grover's search algorithm.}\label{fig16}
\end{figure}

Another important evaluation parameter of the quantum algorithm is circuit depth, which is defined as the longest path from the input to the output, moving forward in time along qubit wires \cite{bib22}. Actually, circuit depth describes the complexity of a quantum circuit in one dimension, and a deeper circuit depth means a more complex quantum circuit. Taking the following quantum circuit as an example, as shown in figure \ref{fig16}, the depth of the circuit is 9.

In \cite{bib22}, the authors have strictly derived the circuit depths for both single-target Grover’s algorithm and the modified Grover's algorithm. Using the analogous method, we expanded them to multi-target situations for Grover’s algorithm, the modified Grover's algorithm and proposed an optimized algorithm as

\begin{equation}\label{eqn-27} 
	dep\left(Grover^\prime s\ algorithm\right)=1+(3M+5)\left\lfloor\frac{\pi}{4}\sqrt{\frac{2^n}{M}}\right\rfloor,
\end{equation}

\begin{equation}\label{eqn-28} 
	dep\left(Long^\prime s\ algorithm\right)=(3M+6)+\left(3M+5\right)\left\lfloor\frac{\pi}{4}\sqrt{\frac{2^n}{M}}-\frac{1}{2}\right\rfloor
\end{equation}
\noindent
and

\begin{equation}\label{eqn-29} 
	dep\left(Optimized\ algorithm\right)=(3M+4)+(3M+3)\left\lfloor\frac{\pi}{4}\sqrt{\frac{2^n}{M}}-\frac{1}{2}\right\rfloor.
\end{equation}

\begin{figure}[htbp]
	\centering
	\includegraphics[width=0.6\textwidth,trim=0 0 0 10,clip]{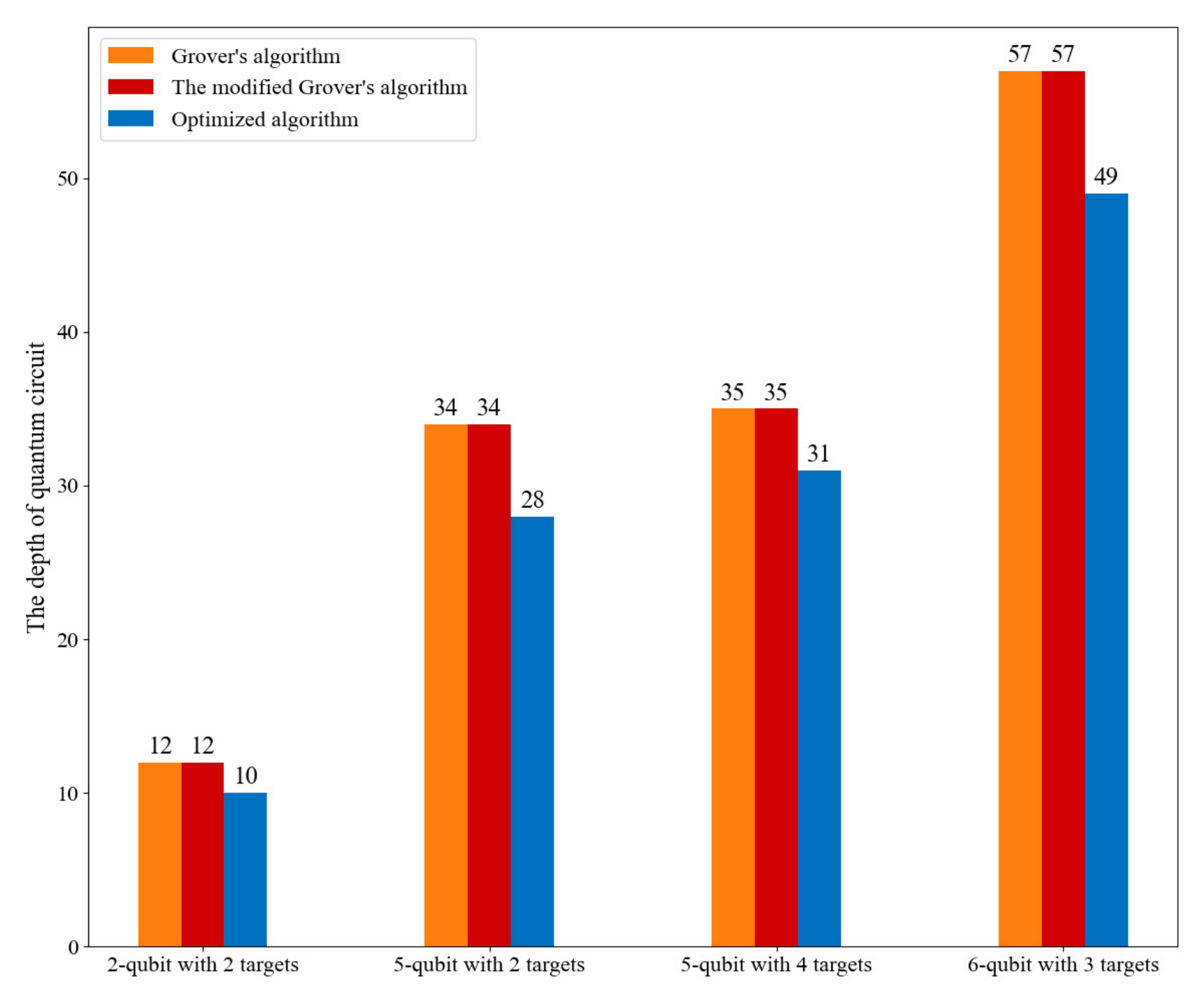}
	\caption{Comparison of the circuit depth for the three different algorithms based on the MindSpore framework.}\label{fig17}
\end{figure}

Where $\left\lfloor A\right\rfloor$ represents the integer part of $A$, $M$ and $n$ are the number of targets and qubits, respectively. 
Figure \ref{fig17} shows a comparison of the circuit depths of the three different algorithms for the above four experimental situations. Again, taking the 5-qubit with 2 targets situation as an example, the
quantum circuit of the proposed optimized algorithm has a circuit depth of 28, while both Grover’s algorithm and the modified Grover's algorithm have circuit depths of 34. This reveals that the circuit depth of the proposed optimized algorithm reduces by about 17.6$\%$ compared with that of Grover’s algorithm and the modified Grover's algorithm. Similarly, the reductions of 16.7$\%$, 11.4$\%$ and 14.0$\%$ can be observed from the other three experimental situations. The above conclusions indicate that the proposed optimized algorithm, by transforming the canonical diffusion operator to a more efficient diffusion operator, can effectively reduce the circuit depth of the quantum circuit.

\section{Conclusion}
In conclusion, we present an optimized exact multi-target search algorithm based on the modified Grover's algorithm by optimizing its diffusion operator to a more efficient diffusion operator, which can solve the multi-target search problem with certainty, while requiring fewer gate counts and shallower circuit depth. After implementing the quantum circuit of the optimized algorithm for 2-qubit with 2 targets, 5-qubit with 2 targets, 5-qubit with 4 targets, and 6-qubit with 3 targets on the MindSpore framework, the results show that, compared with Grover’s algorithm and the modified Grover's algorithm, the optimized algorithm can reduce the quantum gate count by at least 21.1$\%$ and the depth of quantum circuit by at least 11.4$\%$, and maintains a 100$\%$ success probability.

\section*{Acknowledgements}
Thanks for the support provided by MindSpore Community.

\begin{refcontext}[sorting = none]
\printbibliography

@article{bib1,
	title="The computer as a physical system: A microscopic quantum mechanical Hamiltonian model of computers as represented by Turing machines",
	author="Benioff, Paul",
	journal={Journal of statistical physics},
	volume={22},
	pages={563--591},
	year={1980},
	publisher={Springer}
}

@inproceedings{bib2,
	author={Michael A. Nielsen and Isaac L. Chuang},
	year={2011},
	title={Quantum Computation and Quantum Information (10th Anniversary edition)},
	publisher={Cambridge:Cambridge university press}
}

@article{bib3,
	title={A review on quantum search algorithms},
	author={Giri, Pulak Ranjan and Korepin, Vladimir E},
	journal={Quantum Information Processing},
	volume={16},
	pages={1--36},
	year={2017},
	publisher={Springer}
}

@article{bib4,
	title={Quantum theory, the Church--Turing principle and the universal quantum computer},
	author={Deutsch, David},
	journal={Proceedings of the Royal Society of London. A. Mathematical and Physical Sciences},
	volume={400},
	number={1818},
	pages={97--117},
	year={1985},
	publisher={The Royal Society London}
}

@inproceedings{bib5,
	title={Algorithms for quantum computation: discrete logarithms and factoring},
	author={Shor, Peter W},
	booktitle={Proceedings 35th annual symposium on foundations of computer science},
	pages={124--134},
	year={1994},
	organization={Ieee}
}

@inproceedings{bib6,
	title={A fast quantum mechanical algorithm for database search},
	author={Grover, Lov K},
	booktitle={Proceedings of the twenty-eighth annual ACM symposium on Theory of computing},
	pages={212--219},
	year={1996}
}

@article{bib7,
	title={Exactness of the original Grover search algorithm},
	author={Diao, Zijian},
	journal={Physical Review A—Atomic, Molecular, and Optical Physics},
	volume={82},
	number={4},
	pages={044301},
	year={2010},
	publisher={APS}
}

@article{bib8,
	title={Searching a quantum phone book},
	author={Brassard, Gilles},
	journal={Science},
	volume={275},
	number={5300},
	pages={627--628},
	year={1997},
	publisher={American Association for the Advancement of Science}
}

@article{bib9,
	title={Quantum mechanical meet-in-the-middle search algorithm for Triple-DES},
	author={Zhong, PuCha and Bao, WanSu},
	journal={Chinese Science Bulletin},
	volume={55},
	pages={321--325},
	year={2010},
	publisher={Springer}
}

@article{bib10,
	title={Grover search revisited: Application to image pattern matching},
	author={Tezuka, Hiroyuki and Nakaji, Kouhei and Satoh, Takahiko and Yamamoto, Naoki},
	journal={Physical Review A},
	volume={105},
	number={3},
	pages={032440},
	year={2022},
	publisher={APS}
}

@incollection{bib11,
	author	= {Fyrigos, Iosif-Angelos and Dimitrakis, Panagiotis and Ch. Sirakoulis, Georgios},
	title	= {Quantum Computing on Memristor Crossbars},
	booktitle = {Design and Applications of Emerging Computer Systems},
	pages	= {623--647},
	publisher	= {Springer},
	address	=  {Komotini},
	year ={2023},
}

@article{bib12,
	title={Quantum computers can search rapidly by using almost any transformation},
	author={Grover, Lov K},
	journal={Physical Review Letters},
	volume={80},
	number={19},
	pages={43--29},
	year={1998},
	publisher={APS}
}

@article{bib13,
	title={Grover algorithm with zero theoretical failure rate},
	author={Long, Gui-Lu},
	journal={Physical Review A},
	volume={64},
	number={2},
	pages={022307},
	year={2001},
	publisher={APS}
}

@article{bib14,
	title={Arbitrary phases in quantum amplitude amplification},
	author={Hoyer, Peter},
	journal={Physical Review A},
	volume={62},
	number={5},
	pages={052304},
	year={2000},
	publisher={APS}
}

@article{bib15,
	title={Deterministic Grover search with a restricted oracle},
	author={Roy, Tanay and Jiang, Liang and Schuster, David I},
	journal={Physical Review Research},
	volume={4},
	number={2},
	pages={L022013},
	year={2022},
	publisher={APS}
}

@article{bib16,
	title={Better-than-classical Grover search via quantum error detection and suppression},
	author={Pokharel, Bibek and Lidar, Daniel A},
	journal={npj Quantum Information},
	volume={10},
	number={1},
	pages={23},
	year={2024},
	publisher={Nature Publishing Group UK London}
}

@article{bib17,
	title={Quantum Computing (in Chinese)},
	author={Wei, Shijie and Wang, Tao and Ruan, Dong and Long, Guilu},
	journal={Sci Sin Inform},
	volume={47},
	pages={1277--1299},
	year={2017},
}

@article{bib18,
	title={Quantum search with certainty based on modified Grover algorithms: optimum choice of parameters},
	author={Toyama, FM and Van Dijk, Wytse and Nogami, Yukihisa},
	journal={Quantum information processing},
	volume={12},
	pages={1897--1914},
	year={2013},
	publisher={Springer}
}

@article{bib19,
	title={Validating quantum computers using randomized model circuits},
	author={Cross, Andrew W and Bishop, Lev S and Sheldon, Sarah and Nation, Paul D and Gambetta, Jay M},
	journal={Physical Review A},
	volume={100},
	number={3},
	pages={032328},
	year={2019},
	publisher={APS}
}

@article{bib20,
	title={Depth optimization of quantum search algorithms beyond Grover's algorithm},
	author={Zhang, Kun and Korepin, Vladimir E},
	journal={Physical Review A},
	volume={101},
	number={3},
	pages={032346},
	year={2020},
	publisher={APS}
}

@article{bib21,
	title={Subdivided phase oracle for NISQ search algorithms},
	author={Satoh, Takahiko and Ohkura, Yasuhiro and Van Meter, Rodney},
	journal={IEEE Transactions on Quantum Engineering},
	volume={1},
	pages={1--15},
	year={2020},
	publisher={IEEE}
}

@article{bib22,
	title={Distributed exact Grover’s algorithm},
	author={Zhou, Xu and Qiu, Daowen and Luo, Le},
	journal={Frontiers of Physics},
	volume={18},
	number={5},
	pages={51305},
	year={2023},
	publisher={Springer}
}

@article{bib23,
	title={Circuit optimization of Grover quantum search algorithm},
	author={Wu, Xi and Li, Qingyi and Li, Zhiqiang and Yang, Donghan and Yang, Hui and Pan, Wenjie and Perkowski, Marek and Song, Xiaoyu},
	journal={Quantum Information Processing},
	volume={22},
	number={1},
	pages={69},
	year={2023},
	publisher={Springer}
}

@article{bib24,
	title={Novel optimization of quantum search algorithm to minimize complexity},
	author={Kumar, Tarun and Kumar, Dilip and Singh, Gurmohan},
	journal={Chinese Journal of Physics},
	volume={83},
	pages={277--286},
	year={2023},
	publisher={Elsevier}
}

@article{bib25,
	title={Experimental demonstration of deterministic quantum search for multiple marked states without adjusting the oracle},
	author={He, Xin and Zhao, Wen-Tao and Lv, Wang-Chu and Peng, Chen-Hui and Sun, Zhe and Sun, Yong-Nan and Su, Qi-Ping and Yang, Chui-Ping},
	journal={Optics Letters},
	volume={48},
	number={17},
	pages={4428--4431},
	year={2023},
	publisher={Optica Publishing Group}
}

@article{bib26,
	title={Quantum partial search algorithm with smaller oracles for multiple target items},
	author={Li, Dan and Qian, Ling and Zhou, Yu-Qian and Yang, Yu-Guang},
	journal={Quantum Information Processing},
	volume={21},
	number={5},
	pages={160},
	year={2022},
	publisher={Springer}
}

@article{bib27,
	title={Quantum multi-programming for Grover’s search},
	author={Park, Gilchan and Zhang, Kun and Yu, Kwangmin and Korepin, Vladimir},
	journal={Quantum Information Processing},
	volume={22},
	number={1},
	pages={54},
	year={2023},
	publisher={Springer}
}

@article{bib28,
	title={High-dimensional Grover multi-target search algorithm on Cirq},
	author={Acar, Erdi and G{\"u}nd{\"u}z, Sabri and Akp{\i}nar, G{\"u}ven and Y{\i}lmaz, {\.I}hsan},
	journal={The European Physical Journal Plus},
	volume={137},
	number={2},
	pages={1--9},
	year={2022},
	publisher={Springer}
}

@article{bib29,
	title={Search an unsorted database with quantum mechanics},
	author={Long, Guilu and Liu, Yang},
	journal={Frontiers of Computer Science in China},
	volume={1},
	pages={247--271},
	year={2007},
	publisher={Springer}
}

@article{bib31,
	title={Phase matching condition for quantum search with a generalized initial state},
	author={GL Long, X Li, S Yang},
	journal={Physics Letters A},
	volume={294},
	pages={143--152},
	year={2002},
	publisher={Elsevier}
}
\end{refcontext}

\end{document}